\documentstyle[aps,prb]{revtex} 
\begin{document}
\draft
\twocolumn[\hsize\textwidth\columnwidth\hsize\csname
@twocolumnfalse\endcsname

\newcommand{\y}{\'{\i}}

\title{{\em Ab initio} calculations of the dynamical response of copper}
\author{I. Campillo$^{1}$, A. Rubio$^{2}$, and  J. M.
Pitarke$^{1,3}$}
\address{$^1$ Materia Kondentsatuaren Fisika Saila, Zientzi Fakultatea, 
Euskal Herriko Unibertsitatea, 644 Posta kutxatila,\\
48080 Bilbo, Basque Country, Spain\\
$^2$Departamento de F\y sica Te\'orica, Universidad de Valladolid,
E-47011 Valladolid, Spain\\
$^3$Donostia International Physics Center (DIPC) and Centro Mixto CSIC-UPV/EHU}

\date\today

\maketitle

\begin{abstract}

The role of localized $d$-bands in the dynamical response of Cu is
investigated, on the basis of {\em ab initio} pseudopotential calculations.
The density-response function is evaluated in both the random-phase approximation
(RPA) and a time-dependent local-density functional approximation (TDLDA).
Our results indicate that in addition to providing a polarizable
background which lowers the free-electron plasma frequency, $d$-electrons
are responsible, at higher energies and small momenta, for a double-peak structure
in the dynamical structure factor. These results are in agreement with the
experimentally determined optical response of copper.  We also analyze the
dependence of dynamical scattering cross sections on the momentum transfer.
\end{abstract}
]

Noble-metal systems have been the focus of much experimental and theoretical
work in order to get a better understanding of how electronic properties
of delocalized, free-electron-like, electrons are altered by 
the presence of localized $d$-electrons.
Silver is one of the best understood systems, where the free-electron plasma
frequency is strongly renormalized (red-shifted) by the presence of a
polarizable background of $d$-electrons~\cite{Silver}. Unlike silver, copper
presents no decoupling between $sp$ and $d$ orbitals, and a combined description
of these one-electron states is needed to address both structural and electronic
properties of this material. Though in the case of other materials, as
semiconductors, electron-hole interactions (excitonic renormalization) strongly
modify the single-particle optical absorption profile~\cite{excitons}, metals
offer a valuable playground for investigations of dynamical
exchange-correlation effects of interacting many-electron systems within a
quasiparticle picture~\cite{Pines}. Indeed, {\it ab initio} calculations of the
dynamical response of a variety of simple metals, as obtained within the
random-phase-approximation (RPA), successfully account for the experimentally
determined plasmon dispersion relations~\cite{Quong} and scattering
cross sections~\cite{Godby}. Within the same many-body framework, {\it ab
initio} calculations of the electronic stopping power of real solids have also
been reported~\cite{Igor}.

Since the pioneering investigations by Ehrenreich and Philipp\cite{Ehr} on the
optical absorption and reflectivity of Ag an Cu, a variety of measurements of
the optical properties of copper has been reported\cite{Palik}.
Nevertheless, there have been, to the best of our knowledge, no first-principles
calculations of the dielectric response function of Cu that include the full
effects of the crystal lattice. Furthermore, the dynamical density-response
function is well-known to be a key quantity in discussing one-electron
properties in real metals, and it is also of basic importance in the description
of low-dimensional copper systems studied in optical and time-resolved
femtosecond experiments~\cite{Bigot,Angel}.

In this Rapid Communication we report a first-principles evaluation of the
dynamical density-response function of Cu, as computed in
the RPA and a time-dependent extension of local-density functional theory
(TDLDA), after an expansion of $4s^1$ and
$3d^{10}$ one-electron Bloch states in a plane-wave basis with a
kinetic-energy cutoff of $75\,{\rm Ry}$\cite{note1}. Though all-electron mixed
basis schemes, such as the full-potential linearized augmented plane wave
(LAPW) method\cite{Singh}, are expected to be well suited for the description
of the response of localized $d$-electrons, plane-wave pseudopotential
approaches offer a simple and well-defined scenario to describe ground-state
properties and dynamical response functions\cite{Cohen}. This approach has
already been successfully incorporated in the description of inelastic lifetimes
of excited electrons in copper~\cite{Igor2}, and could also be applied to the
study of other noble and transition metals.

The key quantity in our calculations is the dynamical density-response
function $\chi({\bf r},{\bf r}';\omega)$. For periodic crystals we Fourier
transform this function into a matrix $\chi_{{\bf G},{\bf G}'}({\bf
q},\omega)$ which, in the framework of time-dependent density-functional
theory (DFT)\cite{Kohn,Petersilka}, satisfies the matrix equation
\begin{eqnarray}\label{eq1}
&&\chi_{{\bf G},{\bf G}'}({\bf q},\omega)=\chi^0_{{\bf G},{\bf G}'}({\bf
q},\omega)+\sum_{{\bf G}''}\sum_{{\bf G}'''}\chi^0_{{\bf G},{\bf G}''}({\bf
q},\omega)\cr
&&\times\left[v_{{\bf G}''}({\bf q})\delta_{{\bf G}'',{\bf G}'''}+K^{xc}_{{\bf
G}'',{\bf G}'''}({\bf q},\omega)\right]
\chi_{{\bf G}''',{\bf G}'}({\bf q},\omega).
\end{eqnarray}
Here, the wave vector ${\bf q}$ is in the first Brillouin zone (BZ), ${\bf G}$
and ${\bf G}'$ are reciprocal lattice vectors, $v_{\bf G}({\bf q})=4\pi/|{\bf
q}+{\bf G}|^2$ are the Fourier coefficients of the bare Coulomb potential, the
kernel $K^{xc}_{{\bf G},{\bf G}'}({\bf q},\omega)$ accounts for short-range
exchange-correlation effects\cite{Petersilka}, and $\chi^0_{{\bf G},{\bf
G}'}({\bf q},\omega)$ are the Fourier coefficients of the density-response
function of noninteracting Kohn-Sham electrons:
\begin{eqnarray}\label{eq2}
\chi_{{\bf G},{\bf G}'}^0({\bf q},\omega)={1\over \Omega}\sum_{\bf
k}^{BZ}\sum_{n,n'} {f_{{\bf k},n}-f_{{\bf k}+{\bf q},n'}\over E_{{\bf
k},n}-E_{{\bf k}+{\bf q},n'} +\hbar(\omega + {\rm i}\eta)}\cr\cr
\times\langle\phi_{{\bf k},n}|e^{-{\rm i}({\bf q}+{\bf G})\cdot{\bf
r}}|\phi_{{\bf k}+{\bf q},n'}\rangle
\langle\phi_{{\bf k}+{\bf q},n'}|e^{{\rm i}({\bf q}+{\bf G}')\cdot{\bf
r}}|\phi_{{\bf k},n}\rangle,
\end{eqnarray}
where the second sum runs over the band structure for each wave vector
${\bf k}$ in the first BZ, $f_{{\bf k},n}$ are Fermi factors, $\eta$ is a
positive infinitesimal\cite{eta}, and $\Omega$ represents the normalization
volume. The one-particle Bloch states $\phi_{{\bf k},n}({\bf r},\omega)$ and
energies $E_{{\bf k},n}$ are the self-consistent eigenfunctions and eigenvalues
of the Kohn-Sham equations of DFT\cite{Kohn}, which we solve within the so-called
local-density approximation (LDA)~\cite{Ceperley} for exchange and correlation 
effects. The electron-ion
interaction is described by a non-local, norm-conserving, ionic 
pseudopotential~\cite{Troullier} in
a non-separable form, so as to have a better description of
the conduction bands entering Eq. (\ref{eq2}).

The calculations presented below have been found to be well converged for all
frequencies and wave vectors under study, and they have been performed by 
including conduction bands up to a maximum energy of $80\,{\rm eV}$ above the
Fermi level. BZ integrations were performed by sampling on a $10\times
10\times 10$ Monkhorst-Pack mesh\cite{Monk}.  In the RPA the kernel $K^{xc}_{{\bf
G},{\bf G}'}({\bf q},\omega)$ is taken to be zero. In the TDLDA the zero-frequency
kernel, approximated within the LDA by a contact delta function, 
is adiabatically
extended to finite frequencies\cite{Petersilka,Zangwill}. In both RPA and TDLDA,
crystalline local field effects appearing through the dependence of the diagonal
elements of the interacting  response matrix $\chi_{{\bf G},{\bf G}'}({\bf
q},\omega)$ on the off-diagonal elements of the polarizability  $\chi^0_{{\bf
G},{\bf G}'}({\bf q},\omega)$ have been fully included in our
calculations\cite{local_field}.

The properties of the long-wavelength limit (${\bf q}\to 0$) of the dynamical
density-response function are accessible by measurements of the optical
absorption, through the imaginary part of the dielectric response function
$\epsilon_{{\bf G}=0,{\bf G}'=0}({\bf q}=0,\omega)$. On the other hand,
the scattering cross-section for inelastic scattering of either
$X$-rays or fast electrons with finite momentum transfer ${\bf q}+{\bf G}$ is,
within the first Born approximation, proportional to the
dynamical structure factor
\begin{equation}
S({\bf q}+{\bf G},\omega)={2\over v_{\bf G}({\bf q})}{\rm
Im}\left[-\epsilon_{{\bf G},{\bf G}}^{-1}({\bf q},\omega)\right],
\end{equation}
where
\begin{equation}
\epsilon_{{\bf G},{\bf G}'}^{-1}({\bf q},\omega)=\delta_{{\bf G},{\bf G}'}+v_{{\bf
G}'}({\bf q})\chi_{{\bf G},{\bf G}'}({\bf q},\omega).
\end{equation}

Fig. 1 exhibits, by solid lines, our results for both real and imaginary
parts of the $\epsilon_{{\bf G},{\bf G}}({\bf q},\omega)$ dielectric function of
copper, for a small momentum transfer of $|{\bf q}+{\bf
G}|=0.18\,a_0^{-1}$ ($a_0$ is the Bohr radius), together with the optical data
(${\bf q}=0$) of Ref.\onlinecite{Palik} (dashed lines). In this low-${\bf q}$ limit,
both RPA and TDLDA dynamical density-response functions coincide, and the
dielectric function is obtained from the dynamical density-response function of
noninteracting Kohn-Sham electrons\cite{Pines}. Corresponding
values of the so-called energy-loss function
${\rm Im}\left[-\epsilon_{{\bf G},{\bf G}}^{-1}({\bf q},\omega)\right]$ are
presented in Fig. 2, and a comparison between the imaginary parts
of interacting
$\chi_{{\bf G},{\bf G}}({\bf q},\omega)$ and noninteracting $\chi_{{\bf
G},{\bf G}}^0({\bf q},\omega)$ density-response functions is displayed in the
inset of this figure, showing that as the Coulomb interaction is turned on the
oscillator strength is redistributed. Our results, as obtained for a small but
finite momentum transfer, are in excellent agreement with the experimentally
determined dielectric function, both showing a double peak structure in ${\rm
Im}\left[-\epsilon_{{\bf G},{\bf G}}^{-1}({\bf q},\omega)\right]$.

In order to investigate the role of localized
$d$-bands in the dynamical response of copper, we have also 
used an {\it ab initio} pseudopotential with the $3d$ shell assigned to the
core. The result of this calculation, displayed in Fig. 2 by a dotted line,
shows that a combined description of 
both localized $3d^{10}$ and delocalized $4s^1$ electrons
is needed to address the actual electronic response of copper. On the
one hand, the role played in the long-wavelength limit by the Cu
$d$-bands is to provide a polarizable background which lowers the
free-electron plasma frequency by $\sim 2.5\,{\rm eV}$\cite{note2}. We note from
Fig. 1 that near $8.5\,{\rm eV}$ the real part of the dielectric function (${\rm
Re}\,\epsilon$) is zero; however, the imaginary part (${\rm Im}\,\epsilon$)
is not small, due to the existence of interband transitions at these
energies which completely damp the free-electron plasmon. On the other hand,
$d$-bands are also responsible, at higher energies, for a double-peak structure in
the energy-loss function, which stems from a combination of band-structure effects
and the building up of collective modes of $d$-electrons. Since these peaks occur at
energies ($\sim 20\,{\rm eV}$ and
$\sim 30\,{\rm eV}$) where ${\rm Re}\,\epsilon$ is nearly zero (see Fig. 1), they
are in the nature of collective excitations, the small but finite value of ${\rm
Im}\,\epsilon$ at these energies accounting for the width of the peaks.

A better insight onto the origin of the double-hump in the
energy-loss function is achieved from Fig. 3, where the density of states
(DOS) and the joint-density of states (J-DOS) of Cu are
plotted. The high-energy peak present in the J-DOS spectrum at about $25\,{\rm
eV}$, which appears as a result of transitions between $d$-bands at $\sim
2\,{\rm eV}$ below the Fermi level and unoccupied states with energies of
$\sim 23\,{\rm eV}$ above the Fermi level, is responsible for the peak of
electron-hole excitations in ${\rm Im}\,\epsilon$ and ${\rm Im}\,\chi^0$ at
$\omega=25\,{\rm eV}$ (see Fig. 1 and the inset of Fig. 2). Hence, there is a
combination, at high energies, of
$d$-like collective excitations and interband electron-hole transitions, which
results in a prominent double-peak in the loss spectrum.

Now we focus on the dependence of the energy-loss function on the momentum
transfer ${\bf q}+{\bf G}$. As long as the
$3d$ shell of Cu is assigned to the core,
we find a well-defined free-electron plasmon for wave vectors
up to the critical momentum transfer where the plasmon excitation 
enters the continuum of intraband particle-hole excitations. This
free-electron plasmon, which shows a characteristic positive dispersion with
wave vector, is found to be completely damped when a
realistic description of $3d$ orbitals is included in the calculations. At
higher energies and small momenta, $d$-like collective excitations originate a
double-peak structure which presents no dispersion, as shown in Fig. 4. In this
figure the RPA dynamical structure factor for ${\bf G}=0$ is displayed, as obtained
for various values of $q$ along the (100) direction. For larger values of the
momentum transfer ${\bf q}+{\bf G}$, single-particle excitations take over
the collective ones up to the point that above a given cutoff the spectra
is completely dominated by the kinetic-energy term\cite{Pines,Angel}.

In Fig. 5 we show the computed dynamical structure factor for $|{\bf q}+{\bf
G}|=1.91a_0^{-1}$ along the (111) direction, in both RPA
(solid line) and TDLDA (dashed line), 
together with the result of replacing the interacting
$\chi_{{\bf G},{\bf G}}({\bf q},\omega)$ matrix by its noninteracting
counterpart $\chi_{{\bf G},{\bf G}}^0({\bf q},\omega)$ (dotted line). The
noninteracting dynamical structure factor (dotted line) now reproduces the main
features of full RPA and TDLDA calculations. The double peak of Fig.
2, which is in the nature of plasmons, is now replaced by a less pronounced
double-hump originated from single electron-hole excitations. A similar
double-peak has been found in the loss spectra of simple metals\cite{Godby},
which has been understood on the basis of the existence of a gap region for
interband transitions\cite{note3}.
We also note that the effect of short-range
exchange-correlation effects, which are absent in the RPA, is to reduce the
effective electron-electron interaction, thus the dynamical structure factor
being closer in TDLDA than in RPA from the result obtained for noninteracting
Kohn-Sham electrons. The RPA dynamical structure factor of Cu is enhanced by up
to a
$40\%$ by the inclusion, within the TDLDA, of many-body local field corrections.

In summary, we have presented {\it ab initio} pseudopotential calculations of
the dynamical density-response function of Cu, by including $d$-electrons as
part of the valence complex. In the long-wavelength limit (${\bf q}\to 0$),
$d$-bands provide a polarizable background that lowers the free-electron plasma
frequency. $d$-electrons are also responsible for a full damping of this
$s$-like collective excitation and for the appearance of a $d$-like double-peak
structure in the energy-loss function, in agreement with the experimentally
determined optical response of copper. We have analyzed the dependence of the
dynamical structure factor on the momentum transfer, and we have found that, for
values of the momentum transfer over the cutoff wave vector for which collective
excitations enter the continuum of intraband electron-hole pairs, a
less-pronounced double-hump is originated by the existence of interband
electron-hole excitations. Experimental measurements of scattering cross sections in
Cu would be desirable for the investigation of many-body effects, which we have
approximated within RPA and TDLDA.

We thank P. M. Echenique for stimulating
discussions. I.C. and J.M.P. acknowledge partial support by the Basque
Unibertsitate eta Ikerketa Saila and the Spanish Ministerio de Educaci\'on y
Cultura. A.R. acknowledges the hospitality of the Departamento de F\'\i sica de
Materiales, Universidad del Pa\'\i s Vasco, San Sebasti\'an, where part of this
work was carried out.

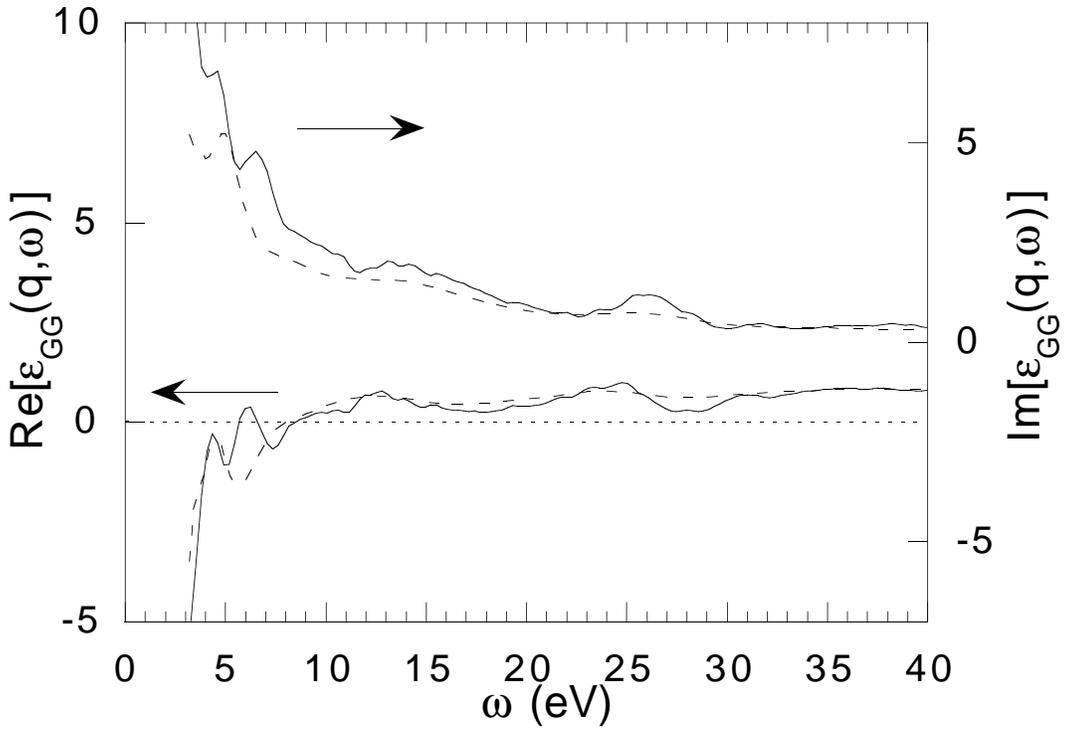
\begin{figure}
\caption{Real and imaginary parts of the $\epsilon_{{\bf G},{\bf G}}({\bf
q},\omega)$ dielectric function of Cu, for ${\bf
q}=0.2(1,0,0)\times(2\pi)/a$ ($a=3.61\,\AA$ is the experimental
lattice constant) and ${\bf G}=0$. Solid and dashed lines represent
our calculations and the optical data of
Ref.\protect\onlinecite{Palik}\protect, respectively.}
\end{figure}

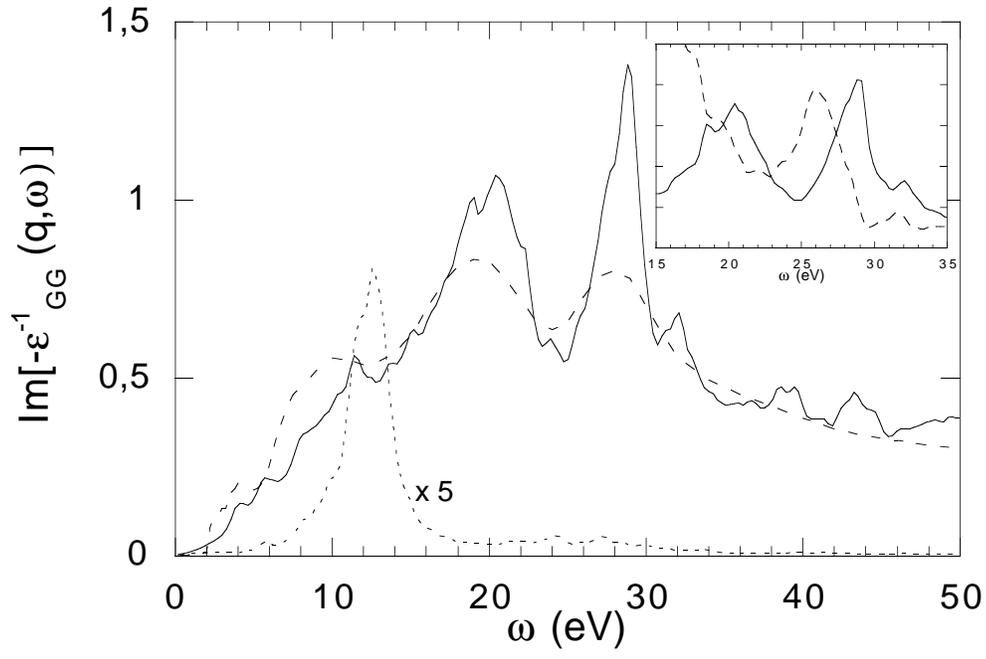
\begin{figure}
\caption{As in Fig. 1, for the energy-loss function of Cu.
The dotted line represents the result of assigning the
$3d$ shell to the core. In the inset the imaginary parts of
interacting (solid line) and noninteracting (dashed line) density-response
functions are compared, for the same values of $\bf q$ and $\bf G$.}
\end{figure}

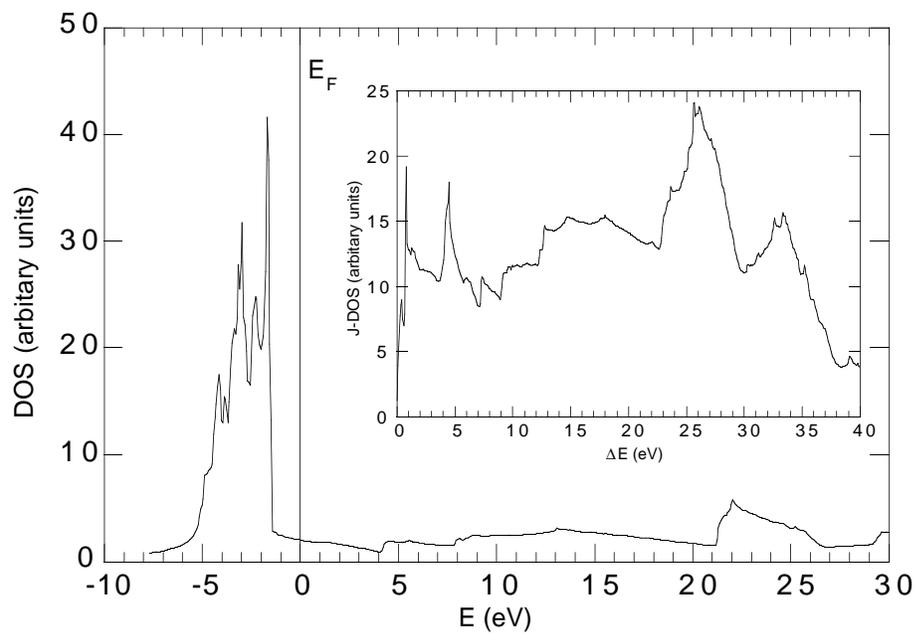
\begin{figure}
\caption{Calculated Density of States (DOS)
and Joint-Density of States (J-DOS) of Cu.
The 25~eV peak in the 
J-DOS mainly corresponds to transitions from the $d$ band to unoccupied states
with energies of $\sim 23\,{\rm eV}$ above the Fermi level. The J-DOS is a first
order approximation to the  optical spectra, when matrix-element renormalization and
Coulomb interactions are neglected. The DOS and the J-DOS were computed
using an interpolataion scheme based on the linear tetrahedron method
with a $20\times 20\time 20$ Mokhrost-Pack mesh.}
\end{figure}

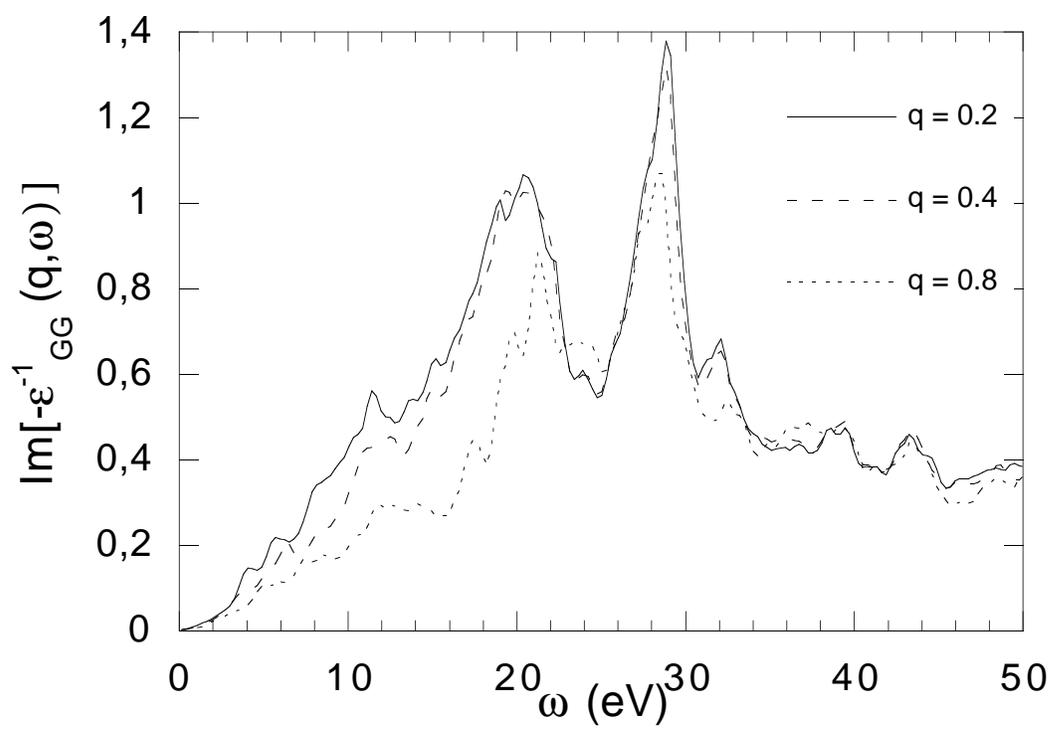
\begin{figure}
\caption{RPA energy-loss function of Cu along the (100) direction,
for various values of the momentum transfer
$|{\bf q}+{\bf G}|$: 0.2, 0.4 and 0.8, in units of
$2\pi /a$ ($a=3.61 \AA$).}
\end{figure}

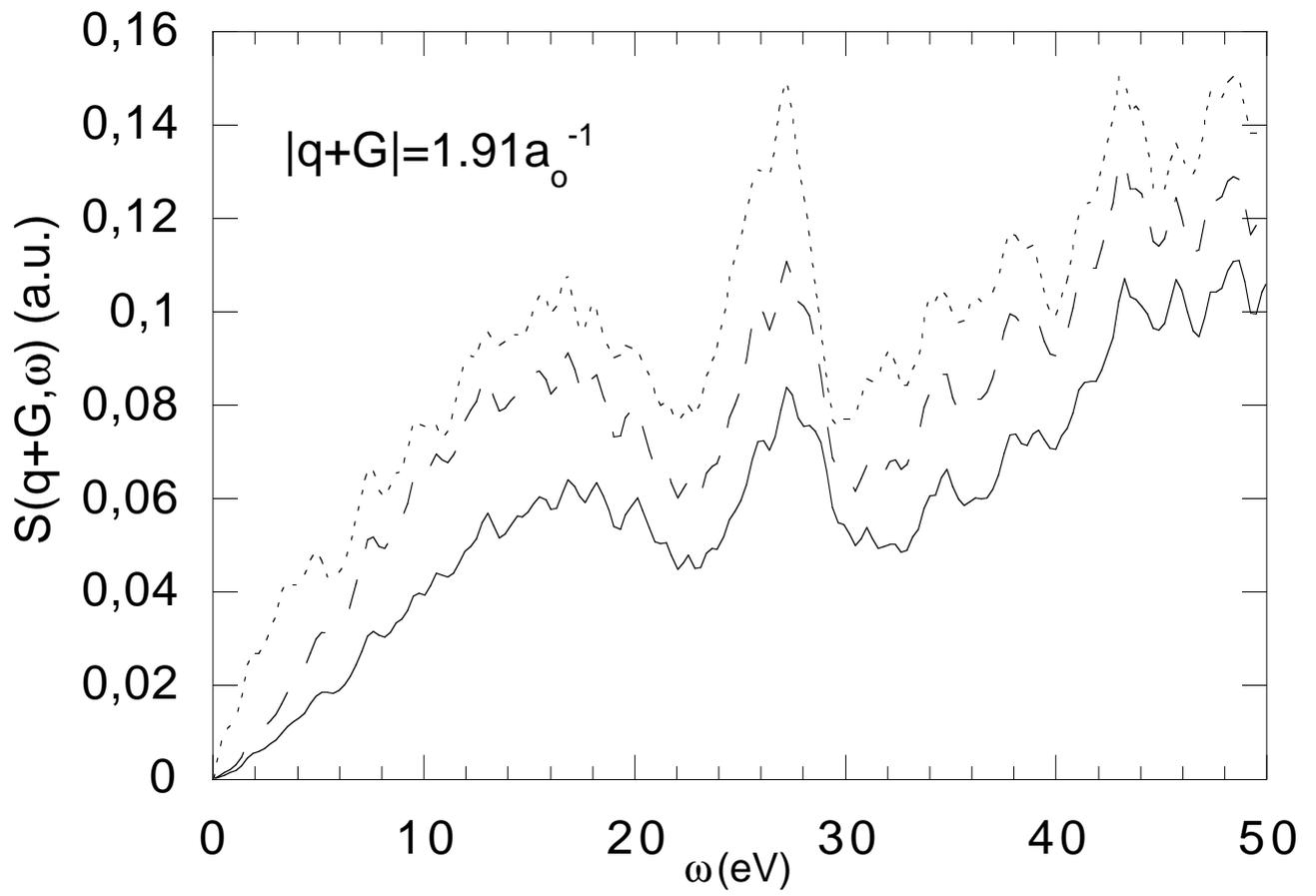
\begin{figure}
\caption{RPA (solid line) and TDLDA (dashed line) dynamical structure factors of
copper, for $|{\bf q}+{\bf G}|=1.91a_0^{-1}$ along the (111) direction
(${\bf q}=0.2(1,1,1)\times(2\pi/a)$ and 
${\bf G}=(1,1,1)\times(2\pi/a)$). The dotted
line represents the result of replacing the interacting density-response matrix
by its noninteracting counterpart.}
\end{figure}


\begin{references}
\bibitem{eta} An important issue is the role of the numerical
damping, $\eta$, introduced in the evaluation 
of Eq. (\ref{eq2}).
We have performed calculations with three different
values of $\eta$: 0.1, 0.2 and 0.4 eV, and we have found no
differences apart from some noise at 0.1 and 0.2 eV.
\bibitem{note2} The plasma frequency of a free-electron gas with the electron
density equal to that of valence ($4s^1$) electrons in Cu is $\omega_{\rm
p}=10.8\,{\rm eV}$.
\end{references}
\end{document}